# Calculation of tunneling rates across a barrier with continuous potential

Sina Khorasani
*School of Electrical Engineering, Sharif University of Technology*
April 9, 2011

Here, approximate, but accurate expressions for calculation of wavefunctions and tunneling rates are obtained using the method of uniform asymptotic expansion [1-3].

Suppose that a classically forbidden potential barrier exists across $a \leq x \leq b$, where the wavenumber $k(x)$ is pure imaginary, or $k^2(x) \leq 0$ with $\text{Im}\{k(x)\} > 0$. If the governing equation is given as

$$\frac{\partial^2}{\partial x^2}\psi(x) + k^2(x)\psi(x) = 0$$

then an approximate forward (left-to-right) tunneling rate from the region $x \leq a$ to $x \geq b$ may be found using the approximate solutions of the Schrödinger's equation based on Airy's functions of the first and second kind. The un-normalized forward-propagating Schrödinger's wavefunction $\psi_+(x)$ is

$$\psi_+(x) = \frac{1}{\sqrt{k(x)}} \sqrt[6]{S_+(x)} Ai\left[+\sqrt[3]{S_+^{\,2}(x)}\right]$$

$$S_+(x) = \frac{3}{2} \int_a^x k(\xi) d\xi$$

Here, $Ai(\cdot)$ is the Airy's function of the first kind. This solution tends to decay for $x \geq a$, but blows up to infinity while approaching $x = b$. To resolve the problem, we select a mid-point $c$ such as $a < c < b$, and for the region $x \leq b$ write down

$$\psi_-(x) = \frac{1}{\sqrt{k(x)}} \sqrt[6]{S_-(x)} Bi\left[+\sqrt[3]{S_-^{\,2}(x)}\right]$$

$$S_-(x) = \frac{3}{2} \int_x^b k(\xi) d\xi$$

Here, $Bi(\cdot)$ is the Airy's function of the second kind. This solution similarly tends to decay for $x \leq b$, but blows up to infinity while approaching $x = a$.

Now, for the particular choice of

$$S_+(c) = \frac{3}{2} \int_a^c k(\xi) d\xi = \frac{3}{2} \int_c^b k(\xi) d\xi = S_-(c) \equiv S$$

we can write down

$$T_{a \to b} \approx T_{c \to b} T_{a \to c} \approx \left|\frac{\psi_-(b)}{\psi_-(c)}\right|^2 \left|\frac{\psi_+(c)}{\psi_+(a)}\right|^2$$

in which

$$\psi_+(a) = \lim_{x \to a} \frac{1}{\sqrt{k(x)}} \sqrt[6]{S_+(x)} Ai\left[+\sqrt[3]{S_+^{\;2}(x)}\right] = \frac{1}{\sqrt[3]{9}\Gamma(\frac{2}{3})\sqrt[6]{\alpha_+}}$$

$$\psi_-(b) = \lim_{x \to b} \frac{1}{\sqrt{k(x)}} \sqrt[6]{S_-(x)} Bi\left[+\sqrt[3]{S_-^{\;2}(x)}\right] = \frac{1}{\sqrt[6]{3}\Gamma(\frac{2}{3})\sqrt[6]{\alpha_-}}$$

$$\psi_+(c) = \frac{1}{\sqrt{k(c)}} \sqrt[6]{S} Ai\left(S^{\frac{2}{3}}\right)$$

$$\psi_-(c) = \frac{1}{\sqrt{k(c)}} \sqrt[6]{S} Bi\left(S^{\frac{2}{3}}\right)$$

where we have made use of the facts [4] that

$$Ai(0) = \frac{1}{\sqrt[3]{9}\Gamma(\frac{2}{3})}$$

$$Bi(0) = \frac{1}{\sqrt[6]{3}\Gamma(\frac{2}{3})}$$

and

$$\lim_{x \to a} \frac{1}{\sqrt{k(x)}} \sqrt[6]{S_+(x)} = \frac{1}{\sqrt[6]{\alpha_+}}$$

$$\alpha_+ = \lim_{x \to a} \frac{k^2(x)}{x-a}$$

$$\lim_{x \to b} \frac{1}{\sqrt{k(x)}} \sqrt[6]{S_-(x)} = \frac{1}{\sqrt[6]{\alpha_-}}$$

$$\alpha_- = \lim_{x \to b} \frac{k^2(x)}{x-b}$$

This helps us to simplify the tunneling rate as

$$T_{a \to b} = 3 \left|\frac{Ai\left(S^{\frac{2}{3}}\right)}{Bi\left(S^{\frac{2}{3}}\right)}\right|^2 \sqrt[3]{\left|\frac{\alpha_+}{\alpha_-}\right|}$$

It is here furthermore possible to simplify the tunneling rates using the appropriate asymptotic expansions of Airy functions as

$$Ai(u) = \frac{1}{2\sqrt{\pi\sqrt{u}}} \exp\left[-\frac{2}{3}\sqrt{u^3}\right], \quad u > 0$$

$$Bi(u) = \frac{1}{\sqrt{\pi\sqrt{u}}} \exp\left[+\frac{2}{3}\sqrt{u^3}\right], \quad u > 0$$

Noting the fact that $\text{Im}\{k(x)\} > 0$ within the classically forbidden region $a \leq x \leq b$, this results in the forward tunneling rate given as

$$T_{a \to b} = \frac{3}{4}\left|\frac{\alpha_+}{\alpha_-}\right|^{\frac{1}{3}} \exp\left[2j \int_a^b k(\xi)d\xi\right]$$

This equation may be considered to be sufficient for most practical purposes. Comparing to the usual expression obtained from WKB expansions, we have

$$T_{a \to b}|_{\text{WKB}} = \exp\left[2j \int_a^b k(\xi)d\xi\right]$$

It is also noteworthy to mention that the approximate solution for the classically allowed region where $k^2(x) \leq 0$ is simply given by

$$\psi(x) = \frac{1}{\sqrt{k(x)}} \sqrt[6]{S(x)} Ai\left[-\sqrt[3]{S^2(x)}\right]$$

so that a uniform solution can be written down as

$$\psi(x) = \frac{1}{\sqrt{k(x)}} \sqrt[6]{S(x)} Ai\left\{\text{sgn}[-k^2(x)]\sqrt[3]{S^2(x)}\right\}$$

with $\text{sgn}(\cdot)$ being the Sign function. Therefore, a general approximate solution to the wave equation is

$$\psi(x) = c_+\psi_+(x) + c_-\psi_-(x)$$
$$\psi_+(x) = \frac{1}{\sqrt{k(x)}} \sqrt[6]{S(x)} Ai\left\{\text{sgn}[-k^2(x)]\sqrt[3]{S^2(x)}\right\}$$
$$\psi_-(x) = \frac{1}{\sqrt{k(x)}} \sqrt[6]{S(x)} Bi\left\{\text{sgn}[-k^2(x)]\sqrt[3]{S^2(x)}\right\}$$

where are $c_\pm$ constants determined by boundary or initial conditions.

**References:**

[1] C. Chester, B. Friedmann, and F. Ursell, An extension of the method of steepest descents, *Proc. Camb. Philos. Soc.* **53**, 599-611 (1957).
[2] N. Bleistein and R. A. Handelsmann, *Asymptotic expansions of integrals*, Holt, Rinehard, and Winston, New York, 1975.
[3] W. P. Schleich, *Quantum Optics in Phase Space*, Wiley-VCH, Berlin, 2001.
[4] M. Abramowitz and I. A. Stegun, eds., *Handbook of Mathematical Functions with Formulas, Graphs, and Mathematical Tables*, Dover, New York, 1965.